\documentclass{emulateapj}
\usepackage{epstopdf}
\usepackage{graphicx}
\usepackage{epsfig}
\usepackage{natbib}

\newcommand\lsim{\mathrel{\rlap{\lower4pt\hbox{\hskip1pt$\sim$}}
        \raise1pt\hbox{$<$}}}
\newcommand\gsim{\mathrel{\rlap{\lower4pt\hbox{\hskip1pt$\sim$}}
        \raise1pt\hbox{$>$}}}

\def\kpch{h^{-1}{\rm kpc}}
\def\Mpch{h^{-1}{\rm Mpc}}

\def\Msun{M_\odot}

\newcommand{\bma}{\begin{math}}
\newcommand{\ema}{\end{math}}
\newcommand{\beq}{\begin{equation}}
\newcommand{\eeq}{\end{equation}}
\newcommand{\beqa}{\begin{eqnarray}}
\newcommand{\eeqa}{\end{eqnarray}}
\newcommand{\bc}{\begin{center}}
\newcommand{\ec}{\end{center}} 
\newcommand{\bit}{\begin{itemize}}
\newcommand{\eit}{\end{itemize}}

\begin{document}

\submitted{\today.} 

\title{Comparison of reionization models: radiative transfer simulations and approximate, semi-numeric models}
\author{Oliver Zahn\altaffilmark{1}, Andrei Mesinger\altaffilmark{2, \dag},
Matthew McQuinn\altaffilmark{3, 4, \ddag}, Hy Trac\altaffilmark{3}, Renyue Cen\altaffilmark{2}, Lars E. Hernquist\altaffilmark{3}
}

\keywords{cosmology: theory -- epoch of reionization -- simulations -- radiative transfer}

\altaffiltext{1}{Berkeley Center for Cosmological Physics, Physics Department and Lawrence Berkeley National Laboratory, University of California, Berkeley, CA 94708}
\altaffiltext{2}{Department of Astrophysical Sciences, Princeton University, Princeton, NJ 08544}
\altaffiltext{3}{Harvard-Smithsonian Center for Astrophysics, 60 Garden Street, Cambridge, MA 02138}
\altaffiltext{4}{Berkeley Astronomy Department, University of California, Berkeley, CA 94720}
\altaffiltext{\dag}{Hubble Fellow}
\altaffiltext{\ddag}{Einstein Fellow}
\email{zahn@berkeley.edu, mesinger@astro.princeton.edu, mmcquinn@berkeley.edu, htrac@cfa.harvard.edu}

\begin{abstract}

We compare the predictions of four different algorithms for the distribution of ionized gas during the Epoch
of Reionization.  These algorithms are all used to run a $100$~Mpc/h simulation of reionization with the same initial conditions.   Two of the algorithms are state-of-the-art ray-tracing radiative transfer codes that use disparate methods to calculate the ionization history. The other two algorithms are fast but more approximate schemes based on iterative application of a smoothing filter to the underlying source and density fields. 
 We compare these algorithms' resulting ionization and $21$~cm fields using several different statistical measures.  The two radiative transfer schemes are in excellent agreement with each other (with the cross-correlation coefficient of the ionization fields $>0.8$ for $k < 10$~h/Mpc) and in good agreement with the analytic schemes ($>0.6$ for $k < 1$~h/Mpc).  When used to predict the 21cm power spectrum at different times during reionization, all ionization algorithms agree with one another at the $10$s of percent level.  This agreement suggests that the different approximations involved in the ray tracing algorithms are sensible and that semi-numerical schemes provide a numerically-inexpensive, yet fairly accurate, description of the reionization process. 

\end{abstract}

\keywords{cosmology: theory -- intergalactic medium -- large scale
structure of universe}

\section{Introduction}\label{sec:intro}

At $z\gsim20$, structures massive enough for the gas to cool and form the first galaxies first broke away from the Hubble flow.  Stars and black holes formed in these structures and ultimately ionized and heated the intergalactic gas.  However, exactly when and how this process occurred are questions that are still unresolved. There are currently two prominent constraints on the reionization era.   First, Ly$\alpha$ forest absorption spectra towards high redshift quasars show a rapidly increasing opacity to Ly$\alpha$ photons by the intergalactic medium (IGM) at $z\gsim6$ \citep[e.g.][]{Fan2006}).  Several studies have interpreted this increase as evidence for the end of reionization.  At the very least, the amount of transmission in the Ly$\alpha$ forest at $z\lesssim6$ indicates that the bulk of reionization occurred at higher redshifts.  Second, measurements of the large-angle cosmic microwave background (CMB) polarization anisotropies suggest that the median redshift of the reionization process was roughly $z\simeq 10.4 \pm 1.4$ \citep{Komatsu2009}.

Taken at face value, these observations favor an extended reionization epoch (see also \citealt{Cen2003, Fan2006, BoltonHaenelt2007b, WyitheCen2007}).  However, the interpretation of quasar absorption spectra is hampered by the large cross section for Ly$\alpha$ absorption, which can lead to complete absorption even if the hydrogen is highly ionized.  This makes the high-redshift Ly$\alpha$ forest spectra difficult to interpret \citep{LidzOF2006, BeckerRS2007}, and the data may even be consistent with reionization completing at redshifts $z<6$ \citep{Lidz2007, Mesinger2009}.
Furthermore, the CMB polarization measurements only offer an {\em integral} constraint on the ionization history \citep{Kogut2003, Page2007}. 

Additional model-dependent constraints on reionization have been derived from other astrophysical probes such as (1) the size of the proximity zone around quasars (\citealt{WyitheLC2005, Fan2006}, but see \citealt{MesingerHC2004, BoltonHaehnelt2007a, Lidz2007, Maselli2007}); (2) a claimed detection of damping wing absorption from neutral IGM in quasar spectra (\citealt{MesingerHaiman2004, MesingerHaiman2007}, but see \citealt{MesingerFurlanetto2008a}); (3) the {\it non}-detection of intergalactic damping wing absorption in a gamma ray burst spectrum \citep{Totani2006,  McQuinn2008}; and (4) the number density and clustering of Ly$\alpha$ emitters \citep{MalhotraRhoads2004, HaimanCen2005, FurlanettoZH2006, Kashikawa2006, McQuinn2007a, MesingerFurlanetto2008b}.

Redshifted 21 cm emission from the hyperfine transition of neutral hydrogen has the potential to provide detailed three-dimensional information about the evolution and morphology of the reionization process \citep[e.g.][]{ZaldarriagaFH2004}.  Several large interferometers will devote significant integration times during the next few years towards detecting this signal.  These efforts include the Mileura Wide Field Array (MWA)\citep{BowmanMH2005}\footnote{http://web.haystack.mit.edu/arrays/MWA/}, 
the Low Frequency Array (LOFAR)\footnote{http://www.lofar.org}, the Giant Metrewave Telescope (GMRT) \citep{Pen2009}, Precision Array for Probing the Epoch of Reionization (PAPER) \citep{Parsons2009}, and eventually the Square Kilometer Array (SKA)\footnote{http://www.skatelescope.org/}. 

Our ability to infer information about the properties of first sources from the 21cm signal, as well as other observations,
hinges on the accuracy of the theoretical modeling of the ionization structure during reionization.  Therefore, a number of groups have developed 3-D radiative transfer (RT) codes (e.g. \citealt{Gnedin2000, Razoumov2002, Ciardi2003, Sokasian2001, Mellema2006, McQuinn2007b, SemelinCB2007, TracCen2007,  AltayCP2008, AubertTeyssier2008, FinlatorOD2009, PetkovaSpringel2009}; see the recent review in \citealt{TracGnedin2009}).  Furthermore, given the large uncertainties in how ionizing photons are produced and escape from galaxies (and also in the dense systems that by the end of reionization are the largest sinks of ionizing photons), a large region of parameter space needs to be modelled in order to interpret observations.  These concerns have prompted several groups to come up with more approximate, but much faster schemes \citep[e.g.][]{Zahn2005, MesingerFurlanetto2007, GeilWyithe2008, Alvarez2009, ChoudhuryHR2009, Thomas2009}. 

Accurate models of the Epoch of Reionization (EoR) must include the evolution of the dark matter, gas, radiation, galaxies, as well as a plethora of feedback processes.  Although, the morphology of reionization may be robust to many of these modeling uncertainties \citep{McQuinn2007b}.  Still, the scope of the parameter space is daunting.  Simulations must resolve the small mass galaxies that are expected dominate the ionizing photon budget (probably corresponding to the atomic cooling threshold, with halo masses of $\sim 10^8\;M_\odot$ at $z\sim$7--10), as well as the photon sinks. 
On the other hand, reionization simulations must also be large enough to statistically sample the distribution of HII regions, which can span tens of comoving megaparsecs in size 
towards the end of reionization
\citep{FurlanettoOh2005, Zahn2005, Zahn2007, MesingerFurlanetto2007, ShinTC2008}.  A few groups have recently come close to achieving this dynamic range (\citealt{Iliev2006b, McQuinn2007b, TracCen2007, ShinTC2008, TracCL2008}; see the recent review in \citealt{TracGnedin2009}).\footnote{Note however that Lyman limit systems (LLSs), which can dominate the absorption of ionizing photons (see for example the appendix of \citealt{FurlanettoOh2005}), are still too small to be resolved by state-of-the-art reionization simulations, and therefore must be included via some analytic prescription.}  In addition, promising analytic alternatives have been suggested \citep[e.g.][]{Furlanetto2004}. The Monte-Carlo implementation of these alternatives in large-volume simulations allows for a description of the non-spherical HII region morphology that can be compared side-by-side with simulations \cite[e.g.][]{Zahn2005, Zahn2007, MesingerFurlanetto2007}.

{\it In this paper, we investigate the convergence of different algorithms for modeling the EoR.}  We compare two different cosmological radiative transfer codes, which follow ionizing radiation from sources identified as halos in an N-body simulation.  Our comparison also includes ionization fields generated with somewhat refined versions of the fast, semi-numerical algorithms mentioned above.  The comparisons are done using the same density field, and N-body halo field (where applicable).  We test the codes on the full reionization problem (at least to the extent that it is currently simulated) in order to study the numerical convergence of the predicted 21 cm signal. Our approach is different but complimentary to \citet{Iliev2006a, Iliev2009} who compared several RT codes using a number of tests problems. Both approaches are necessary for studying numerical convergence.

The structure of this paper is as follows. In Section \ref{sec:simulations}, we present the N-body simulation we employ, as well as the two different radiative transfer schemes that are run in post-processing on top of the N-body simulation.  In Section \ref{sec:seminumeric}, we describe the approximate schemes. Section \ref{sec:statcomp} investigates the spatial agreement between the ionization fields of these algorithms.  Section \ref{sec:21cm} compares the predictions of these schemes for the 21cm power spectrum. In the Appendix, we present further discussion and tests of the semi-numerical schemes.

The simulations and analytic calculations presented here are based on a $\Lambda CDM$ cosmology with the following parameter values: $\sigma_8=0.82, h=0.7, \Omega_{\rm m}=0.28, \Omega_{\rm b}=0.046, n_s=0.96$, consistent with latest constraints from WMAP \citep{Komatsu2009}.

\section{Test Problem}
\label{sec:simulations}

In this paper, the radiative transfer and semi-numeric calculations were performed by post-processing input density fields from \citet{TracCL2008}. A high-resolution N-body simulation with $3072^3$ dark matter particles was used to evolve the matter distribution in a periodic box of comoving volume $(100\ \Mpch)^3$. We chose to work with 82 snapshots of the matter density field (every 10 million years from redshift $z=27$ to $z=6$), gridded onto a $256^3$ Cartesian grid for the reionization algorithms. The baryons were assumed to trace the dark matter down to the grid cell spacing $\Delta x=390\ \kpch$, which is close to the anticipated physical smoothing scale for the intergalactic gas (the Jeans length for $10^4$~K gas is equal to $370\ \kpch$ at the mean density at $z=6$).

Radiation sources embedded in the large-scale structure of the matter distribution were modeled using halos catalogued in the N-body simulation. Dark matter halos were located on the fly using a friends-of-friends algorithm, with a linking length of $b=0.2$ times the mean interparticle spacing. Halos with virial temperatures above the atomic cooling limit ($T\gtrsim10^4$ K; or $M\gtrsim10^8\Msun$ for the relevant redshifts) were located with a minimum of $\sim40$ particles ($M=1.0\times10^8\Msun$), and half of this collapsed mass budget is resolved with $>400$ particles per halo. The first source appeared at $z\sim27$ within the $(100\ \Mpch)^3$ comoving volume. By $z=6$, there were $>6$ million sources and, when binned on a $256^3$ Cartesian grid, the clustered sources occupied $\sim17\%$ of the grid.

\section{Radiative Transfer Algorithms}
\label{sec:radiativetransfer}

This section describes the two radiative transfer codes that are performed on the test problem described in Section \ref{sec:simulations}:  the McQuinn et al. code (Section \ref{subsec:rt_mcquinn}) and the Trac \& Cen code (Section \ref{subsec:rt_trac}).

\subsection{McQuinn et al.}
\label{subsec:rt_mcquinn}

The algorithm presented in \citet{McQuinn2007b} is an improvement of the \citet{Sokasian2001} ray tracing code.  It is a more approximate ray tracing scheme than \citet{TracCen2007}, but has more modest CPU and memory requirements.  This code was specifically designed to perform ray tracing on the millions of sources required to study the EoR and was the first ray tracing code to achieve this feat \citep{McQuinn2007b}.\footnote{Although, HI cooling mass halos were unresolved in the N-body simulation used in this study and were instead included in post processing prior to ray tracing.}  In addition, this code has enabled the first parameter space studies of this epoch \citep{McQuinn2007b}.

This code sends out $768$ rays from every source between every time step, and rays travel within that time step until their photons are expended (i.e., an infinite speed of light).  It uses the HEALPix-based adaptive ray-splitting scheme of \citet{AbelWandelt2002}, such that rays split as they travel to ensure that a minimum number of rays reach each cell.  For the simulations in this paper, each cell receives at least $2.1$ rays from each source when averaged over HEALPix orientations.  The numbers $768$ (corresponding to HEALPix level $3$) and $2.1$ were calibrated with convergence tests in the Appendix of \citet{McQuinn2007b} and were the values that this code has been run with for large-scale simulations of cosmological reionization.   The order that rays are cast in this code is randomized over both the sources and directions for each time step.   Finally, in the test problem in this paper, each time step is taken to be $\Delta t =10$~Myr, which corresponds to the time between stored snapshots of the density field.  Comparable values for $\Delta t$ have been used in published studies with this code \citep{McQuinn2007a, McQuinn2007b}.  

The McQuinn et al. radiative transfer algorithm for hydrogen reionization works as follows.  At the beginning of the time step, the ionization fraction in each cell is corrected for recombinations that occur during that time step.  Next, rays are cast.  Each ray carries a set number of photons and when a ray reaches a cell that contains neutral gas, it deposits all of the photons into ionizations.  The approximation that all the photons are absorbed at the ionization front is motivated by the short mean free path of ionizing photons in a neutral medium for a physically motivated source spectrum during hydrogen reionization (i.e., a mean free path of $10$s of kpc is expected, which is much smaller than the cell size).  When the path to the edge of an ionized region becomes comparable to the box size, this algorithm slows significantly because it has to follow each ray to the front edge each timestep.
The solution to the ionization field that this algorithm obtains technically depends on the order rays are cast from each cell, but convergence studies have demonstrated that in practice a nearly identical solution is reached with different ray orderings \citep{McQuinn2007b}.  After rays have ionized a cell, it is assumed to be in photoionization equilibrium with the photoionizing background.

Finally, as we mentioned previously, this algorithm assumes an infinite speed of light.  This approximation is the least valid at the end of reionization when the HII regions are the largest.  
 Because of this approximation, the McQuinn et al. algorithm produces several per cent larger ionization fractions close to the end of reionization, compared with the other radiative transfer algorithm in this comparison by \citet{TracCen2007}.  This issue could be alleviated in higher resolution simulations than performed here where Lyman limit absorption systems are resolved, which will limit the mean free path.

\subsection{Trac \& Cen}
\label{subsec:rt_trac}

The \citet{TracCen2007} radiative transfer algorithm follows the propagation of ionizing photons using an adaptive ray tracing technique, featuring several novel approaches that distinguish it from other previous ray tracing algorithms. The major characteristics include conservation of photons, a time-dependent solution that is causal, and computational scaling that is independent of the number of sources. Here we provide a basic description of the algorithm and highlight differences with other approaches.

As a basis, the adaptive ray-splitting scheme of \cite{AbelWandelt2002} is used to efficiently improve spatial resolution. For a given source, a preset number of base rays are cast and they travel a short distance before each splits into 4 daughter rays. Successive generations of splitting continue downstream such that the angular resolution continually increases farther away from sources. This procedure ensures that a minimum number of rays intersect each grid cell element of the density field within the ionized region. For this comparison study, the number of base rays was set at 192 and the minimum number of rays intersecting each grid cell was set at 2.

One major feature of the algorithm is that new source rays are cast for every radiative transfer time step, which is set equal to the light-crossing time, defined as the length of a grid cell divided by the speed of light. Correspondingly, all rays are advanced a maximum distance (when the medium is optically thin) equal to a grid cell length in each time step. This synchronization ensures a causal solution, a characteristic usually not shared by other ray tracing algorithms where rays from multiple sources are treated independently. Generally, this approach is costly because the light-crossing time at high resolution can be quite small compared to the duration of reionization. However, \citet{PawlikSchaye2008} have pointed out that when the radiation filling factor is close to unity near the end of reionization, the light-crossing time stepping can actually be computationally faster than the infinite speed of light approach.

Another major feature of the algorithm is an adaptive ray-merging scheme designed to circumvent the ${\cal O}(N^2)$ scaling in the multi-source problem. In the brute-force approach, every source must emit roughly as many rays as there are density grid cells when the radiation filling factor is close to unity. For a $256^3$ grid with close to 3 million source cells, a computationally infeasible number of 50 trillion rays would be required in total. \citet{TracCen2007} adaptively restrict the maximum number of rays entering and exiting a given cell by merging near-parallel rays from multiple sources. For this comparison study, approximately 100 rays per cell and as many as 2 billion rays to propagate within a time step were required by the end of reionization. One disadvantage is that a large amount of memory is still required to store the necessary rays. In the current version of the implementation, each ray requires 44 bytes to store intrinsic variables plus another 8 bytes per frequency bin to keep track of the frequency and photon count.

For each radiative transfer time step, the photoionization rates are calculated on the same $256^3$ grid using the incident radiation flux from the rays. The equations for the ionization evolution are then solved using a non-equilibrium solver where the time integration involves a stable implicit scheme. For this comparison study, only photoionization at one frequency is considered.  However, the radiative transfer algorithm can also handle photo-heating and multiple frequencies in a cosmological hydrodynamic simulation \citep{TracCL2008}.

\section{Semi-numeric modeling}\label{sec:seminumeric}

Our comparisons also include a semi-numerical algorithm for generating the ionization field.  These techniques are motivated by the intuition that sources ionize the regions immediately around them and that the clustering of sources drives the structure of reionization.  The radiative transfer algorithms described previously can require large computer clusters and can take days to process a single redshift output.  However, for many cosmological reionization applications, it might be useful to sacrifice some accuracy for an increase in speed and dynamic range.  This is potentially quite useful for many applications, such as in simulating the signal for the upcoming 21cm surveys.  The effective resolution of all planned instruments, including the SKA, is much worse than the resolution of the typical RT scheme.  Additionally the field of view (FoV) of some of these instruments will be enormous (e.g. 5000$\times$5000 Mpc$^2$ at $z\sim8$ for MWA), far out of reach of conventional RT algorithms hoping to include sources down to the atomic hydrogen cooling threshold\footnote{It is not clear how useful simulating the FoV of these instruments would be, given that foreground subtraction will likely render most of the transverse modes unusable for studying the EoR \citep[e.g.][]{MoralesWyithe2009}.
  However, in order to get proper sampling of the line-of-sight (LOS) power, one still needs many EoR realizations, as otherwise the LOS modes will add coherently. It would be most desirable to correctly include cosmic variance in these realizations, either through a very large box/dynamic range, or through many smaller boxes where long wavelengths modes have been added-in ``by hand''.}.
  Also, very little is presently known about the properties of the underlying sources, translating to an enormous parameter space in need of exploration.  Thus, a ``relatively'' accurate yet fast ionization algorithm can be an invaluable tool in reionization studies.  Below we detail the semi-numerical algorithm we use, capable of generating large-scale 3D ionization fields\footnote{``Ionization fields'' is synonymous with ``distribution of HII regions'' in this paper.} in a matter of minutes on a single processor (see \citealt{Zahn2005, Zahn2007, MesingerFurlanetto2007, GeilWyithe2008, Alvarez2009, ChoudhuryHR2009, Thomas2009} for related/alternate semi-numerical ionization algorithms).

Our semi-numerical algorithm is based on the excursion-set formalism for modeling reionization developed by \citet{Furlanetto2004}. The foundation of this approach is to require that the number of ionizing photons inside a region be larger than the number of hydrogen atoms it contains.  Then ionized regions are identified via an excursion-set approach, starting at large scales and progressing to small scales, analogous to the derivation of the Press-Schechter mass functions \citep{Bond1991, LaceyCole1993}.

We extend the \citet{Furlanetto2004} scheme to 3D realizations by requiring that an ionized simulation cell at coordinate ${\bf x}$ meet the criteria (c.f. \citealt{Zahn2005, MesingerFurlanetto2007}).

\begin{equation}
\label{eq:HII_barrier}
f_{\rm coll}({\bf x}, z, R) \geq \zeta^{-1} ~ ,
\end{equation}

\noindent where $\zeta$ is some efficiency parameter and $f_{\rm coll}$ is the fraction of mass residing in collapsed halos inside a sphere of mass $M=4/3 \pi R^3 \bar{\rho} [1+\langle \delta_{\rm NL} \rangle_R]$, with mean physical overdensity $\langle \delta_{\rm NL} \rangle_R$, centered on Eulerian coordinate ${\bf x}$, at redshift $z$.
  Following the excursion-set approach \citep{Bond1991, LaceyCole1993}, we decrease the filter scale $R$, starting from some maximum value $R_{\rm max}$\footnote{ $R_{\rm max}$ can in principle can be set to correspond to the mean free path of ionizing photons \citep{FurlanettoOh2005, ChoudhuryHR2009}.  As the mean value and spatial fluctuations of the mean free path are poorly known at present, and as they are not fully included in the numerical simulations, for purposes of comparison we use $R_{\rm max}\sim$ $89$ Mpc, which corresponds to the entire simulation box.  Note however, that in the regime of interest when $R_{\rm max} \gg$ the typical HII region size, the precise value of $R_{\rm max}$ does not affect ionization morphology.}, and progressing down to the cell size, $R_{\rm cell}$, with logarithmic filter spacing of $R_{\rm next} = 0.9 R_{\rm prev}$.  If at any filter scale the criterion in eq. \ref{eq:HII_barrier} is met, this cell is flagged as ionized. Note that we only flag the central filter cell at {\bf x} as ionized (as in \citealt{Zahn2005}) instead of ``painting'' the entire sphere with radius $R$ as ionized (ala \citealt{MesingerFurlanetto2007}), as the former approach is faster and thus more versatile.

 Unlike in \citet{Zahn2005} and \citet{MesingerFurlanetto2007}, we take into account partially-ionized cells by setting the cell's ionized fraction to $\zeta f_{\rm coll}({\bf x}, z, R_{\rm cell})$ at the last filter step for those cells which are not fully ionized.  More specifically, we use the unfiltered value from the cubical simulation cell instead of filtering it with a spherical filter on the cell size, in order to remove aliasing effects from using an analytic filter form. A two-phased medium (containing only fully ionized and fully neutral cells) is a good approximation for stellar sources and simulation resolution we study (see the Appendix).  Nevertheless, partially ionized cells become much more common as the cell size is increased.  Our prescription accounts for cells which have been partially ionized by the sources they contain (see also \citealt{GeilWyithe2008}).

We present two main variations of this semi-numerical ionization algorithm, distinguished by their method of computing $f_{\rm coll}$ and choice of filter:
\begin{itemize}
\item {\it Fast Fourier Radiative Transfer (FFRT)}:   This variant computes $f_{\rm coll}$ using the conditional Press-Schechter (PS) formalism \citep{LaceyCole1993, SomervilleKolatt1999} directly on the {\it evolved} (i.e., non-linear) density field. Although the collapse criterion for the conditional Press-Schechter was derived using the linear density field, we find that using the evolved density from simulations is a better match to the RT ionization fields. This is to be expected because the evolution leads to nonlinear shifts with respect to the linear field. Therefore, this scheme bypasses the need to resolve halos, allowing for a much larger dynamical range. The other modification compared to earlier semi-numeric algorithms is the use of a sharp k-space filter instead of spherical top-hat\footnote{ Although the spherical top-hat is more physically intuitive (ionizing photons in a sphere are compared to neutral hydrogen atoms in the same sphere), the sharp k-space filter results in a somewhat better correlation of the ionization and density fields in the FFRT scheme, which is of importance in predicting the 21cm signal (see Fig. \ref{fig:21cm_power} and associated discussion).  Another advantage of this new filter is that $\zeta f_{\rm coll}$ will be to equal the global mean ionized fraction, in the limit of large box sizes (see Appendix of \citealt{Zahn2007}). The top-hat filter performs much worse in this respect.  Therefore, the redshift evolution of reionization is more accurately predicted using a sharp k-space filter than using a top-hat filter.
}. 

\item {\it Fast Fourier Radiative Transfer-Sources (FFRT-S)}:  This variant computes $f_{\rm coll}$ by filtering the N-body halo field (more precisely the star-formation rate field).  This scheme therefore requires a discrete source field, obtained either through N-body simulations \citep{Zahn2007} or through another semi-numerical scheme \citep{MesingerFurlanetto2007}.  This algorithm by default uses a top-hat filter, in order to avoid ``ringing'' artifacts in the ionization field which can result from the oscillatory configuration-space filter response of the sharp k-space\footnote{In practice, we do not find evidence for such artifacts for the regimes studied here.  However, they should become more common in the regime where sources are rare.  Therefore our fiducial scheme operating on discrete sources uses a top-hat filter.}.
\end{itemize}

In setting the partial ionizations in the FFRT scheme, we also account for Poisson fluctuations in the halo number, when the mean collapse fraction obtained through conditional Press-Schechter\footnote{Specifically we use the ``hybrid'' form of \citet{BarkanaLoeb2004}, which includes the additional factor of the ratio of the PS and Sheth-Tormen \citep{ShethTormen1999} collapse fractions at mean density.  \citet{BarkanaLoeb2004} find that this change in the overall normalization yields mass functions which better compare to N-body simulations.  Note that, aside from computing the partially ionized cells, the difference between this hybrid form and PS can be absorbed in the $\zeta$ efficiency parameter.}
 becomes small, $f_{\rm coll}({\bf x}, z, R_{\rm cell}) [1+\delta_{\rm cell}({\bf x}, z, R_{\rm cell})] \bar{M}_{\rm cell} < 50M_{\rm min}$, where $\bar{M}_{\rm cell}$ is the mean, total (DM+baryonic) mass corresponding to a cell, $\delta_{\rm cell}$ is the fractional overdensity in the cell's mass, and our faintest ionizing sources correspond to halo masses of $M_{\rm min} \sim 10^8 M_\odot$ in this application.  We make the simplifying assumption that the sub-grid sources are all of the same mass, $M_{\rm min}$ (likely a decent approximation for partially ionized cells, since the mass function is fairly steep in this regime).  We sample a Poisson distribution with mean $f_{\rm coll} \bar{M}_{\rm cell} (1+\delta_{\rm cell}) / M_{\rm min}$ to obtain the number of sub-grid sources, $N_{\rm min}$.  We then compute that cell's collapse fraction, $f_{\rm coll~sampled} = N_{\rm min}  M_{\rm min} / \bar{M}_{\rm cell} / (1+\delta_{\rm cell})$. Poisson fluctuations are found to be important when the cell size increases to $\geq 1$ Mpc (see Fig. \ref{fig:PDFs} and associated discussion).  Note however that this feature should be left as an option, depending on the particular application.  Although improving the statistics of the ionization fields at a fixed redshift, such {\it stochastic} fluctuations in the collapsed mass and ionization fields make it impossible to {\it deterministically} track the redshift evolution of a particular realization.  Halos would effectively ``pop in-and-out of'' existence from one redshift/time step to the next \citep[e.g.][]{Santos2009}.

This semi-numerical scheme can be extended to contain additional physics such as spatially inhomogeneous recombinations \citep{ChoudhuryHR2009}, a mass dependent ionization efficiency  \citep{FurlanettoMH2006}, and radiative feedback \citep{GeilWyithe2008}.  Since these processes are also poorly understood and the RT schemes we compare against do not take them into account, we will limit our refinements in this work to the simplest case that ignores feedback and inhomogenous recombinations.


\section{Comparison of Ionization Fields}\label{sec:statcomp}

\begin{figure*}
\bc
\includegraphics[width=17cm]{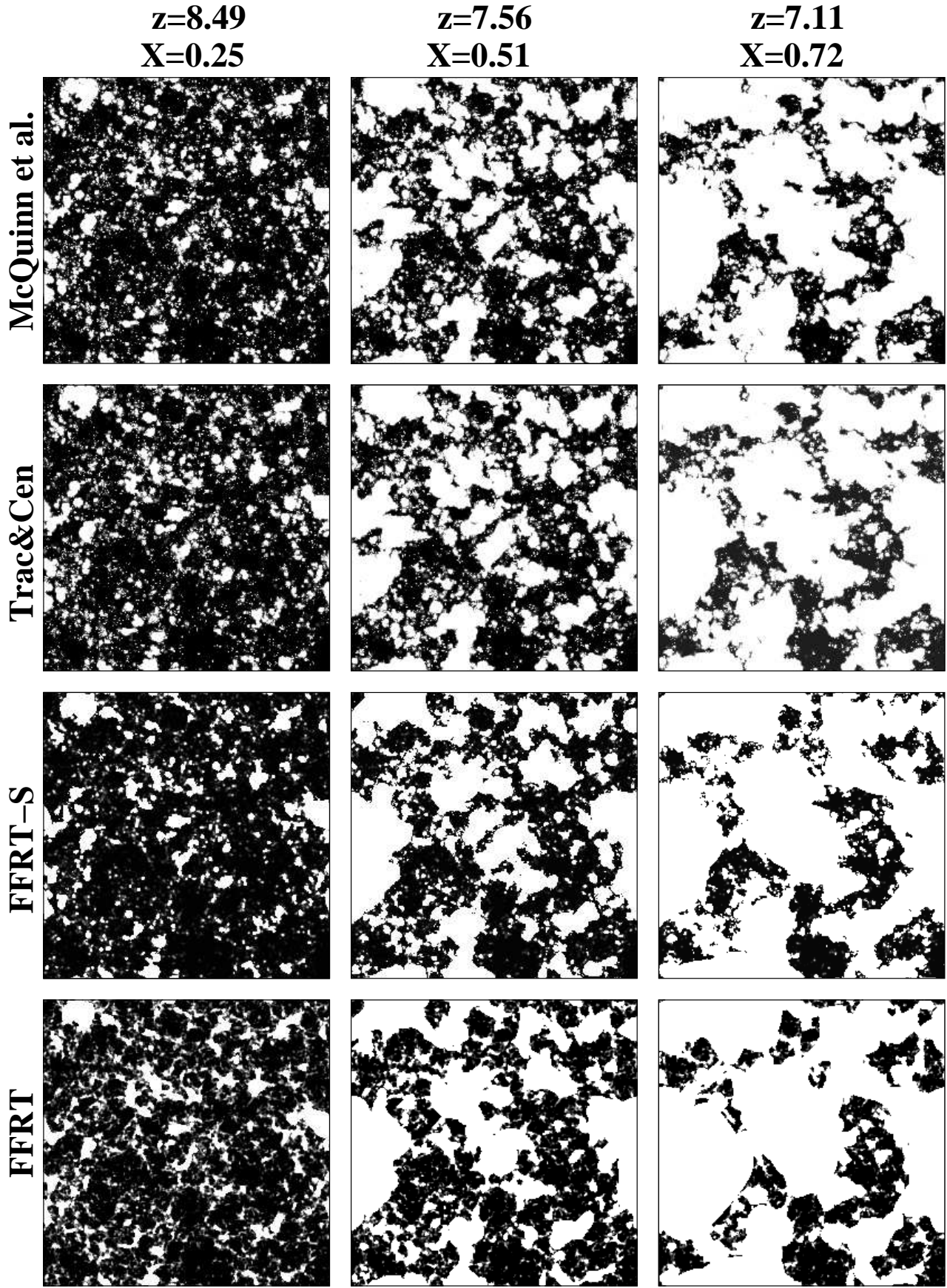}
\caption{Comparison of ionization fields generated from four schemes: McQuinn et al., Trac \& Cen, FFRT-S, and FFRT. The maps are from the same slice (100 Mpc/h by 100 Mpc/h with depth of 0.4 Mpc/h) through the simulation box.}
\label{fig:xfields}
\ec
\end{figure*}

This section compares the ionization fields generated from the four schemes: \citet{McQuinn2007b}, \citet{TracCen2007}, FFRT, and FFRT-S. Figure \ref{fig:xfields} shows ionization maps for three different volume-weighted ionization fractions, $x_{i,V}=0.25, 0.51$, and $0.72$.  The maps are from the same slice, spanning 100 Mpc/h by 100 Mpc/h with a depth of 0.4 Mpc/h, through the simulation box. In all of the simulations, the reionization process appears qualitatively similar. In the early stages, HII regions typically contain relatively few galaxies and are spatially clustered similar to the sources. The HII regions expand and merge with one another as reionization progresses. By the time the process is half completed, the ionization structures appear much more connected, resembling the large-scale-structure of the universe. The largest HII regions now contain tens of thousands of galaxies. In the final stages, the ionization fronts quickly expand into the remaining neutral regions until there is complete overlap.

The maps from the two RT simulations show the closest agreement, with very similar ionization structures at all times. The ionization fronts are similarly resolved, such that even small neutral regions between converging fronts are well preserved. Of the two semi-numeric methods, the FFRT-S produces maps that are in closer agreement with the RT results.  Although reproducing the HII morphology on moderate to large scales, the semi-numeric algorithms tend to generate more connected HII regions.  Such differences are understandable, as the semi-numeric models do not include the physics needed to exactly capture the position and structure of the ionization fronts. In the rest of this section, we compare several statistics to quantify the reionization morphology and to check the degree of convergence between the different schemes.

\subsection{Bubble size distribution}

\begin{figure}
\bc
\includegraphics[width=9cm]{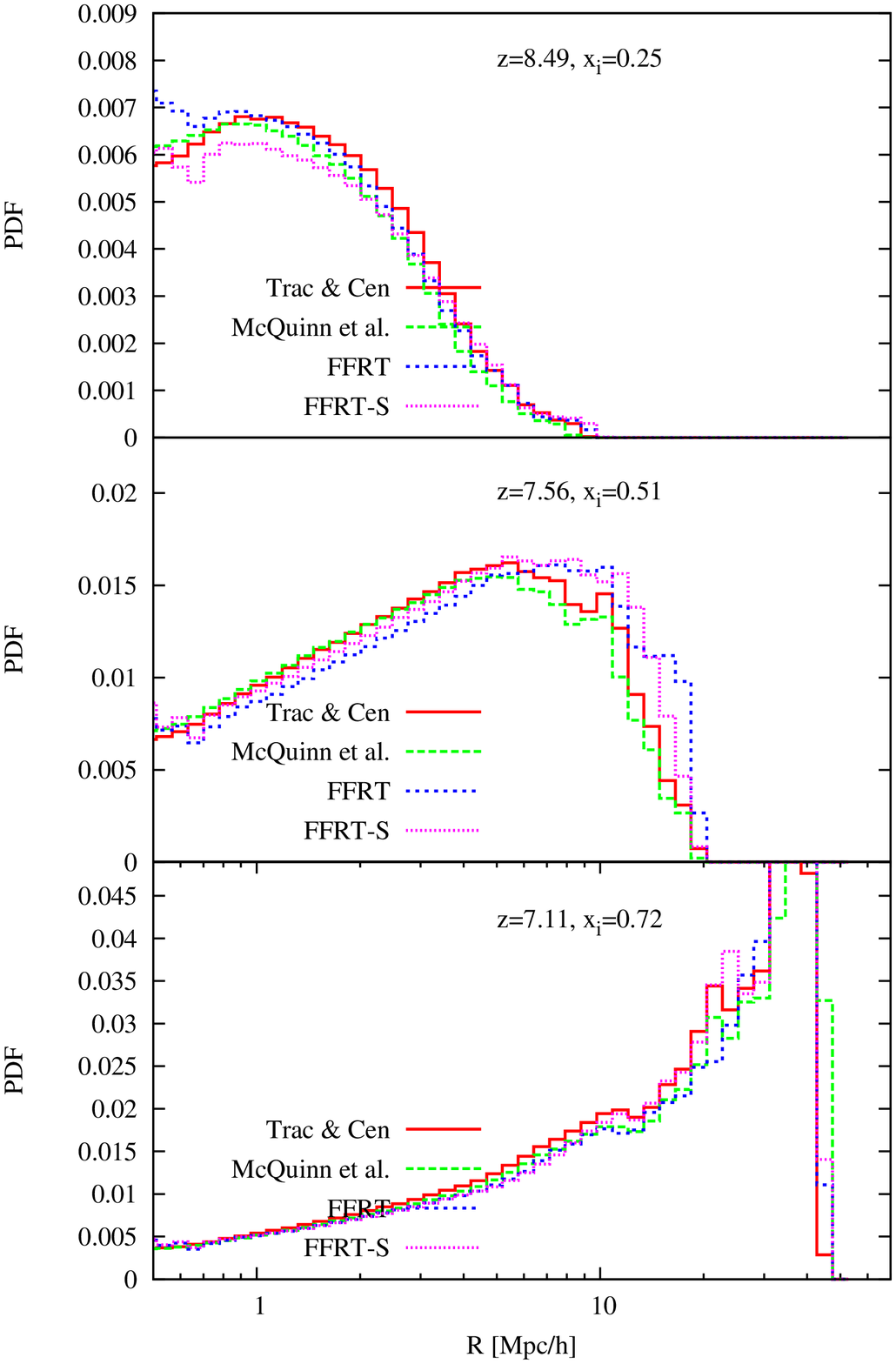}
\caption{The ionized bubble size distribution for the four schemes: Trac \& Cen (red solid), McQuinn et al.~(green dashed), FFRT (blue short-dashed), and FFRT-S (magenta dotted). Plotted is the likelihood that a given ionized patch is part of a region with size $R$. The ionized bubbles get noticeably larger with increasing ionization fraction with all the methods and the trends and peak scales agree among all the methods.}
\label{fig:pdf_xi}
\ec
\end{figure}

Figure \ref{fig:pdf_xi} shows the ionized bubble size distribution, defined as the probability that a randomly chosen ionized cell from the simulation volume is part of an ionized region of radius $R$. An ionized cell is assigned to be in a bubble of radius $R$, where $R$ is set equal to the largest sphere surrounding the cell whose mean enclosed ionized fraction is $0.9$ \citep[see][for more discussion]{Zahn2007}. This definition is convenient because it is similar to that which is used in excursion set reionization models \citep{Furlanetto2004}. In addition, it has the advantage of being quickly computable with Fast Fourier Transforms. The bubble size distribution has a characteristic peak size that shifts toward larger scales as HII regions grow and reionization progresses.

The distributions from the four schemes are generally in good agreement, as expected from the visual inspection of the ionization maps. The two RT simulations show very good agreement on most scales and at all times. Some minor differences are found in the sizes of the largest and rarest of HII regions. At early times, the FFRT algorithm produces more small bubbles while the FFRT-S produces less relative to the RT simulations.  These differences might be due to the disparate default filters.  The top-hat filter of the FFRT-S scheme, operating on relatively few discrete sources early in reionization, tends to produce less connected HII regions (and hence more small bubbles according to the definition used here).  On the other hand, the sharp k-space filter used in FFRT can include ionizing photons from more distant sources through its non-vanishing real space filter response, which should result in more connectivity of HII regions. Later in reionization when sources are more abundant, both semi-numeric schemes tend to produce slightly larger bubbles because of more connectivity between ionized regions.

\subsection{Power spectra}

\begin{figure}[t]
\bc
\includegraphics[width=9cm]{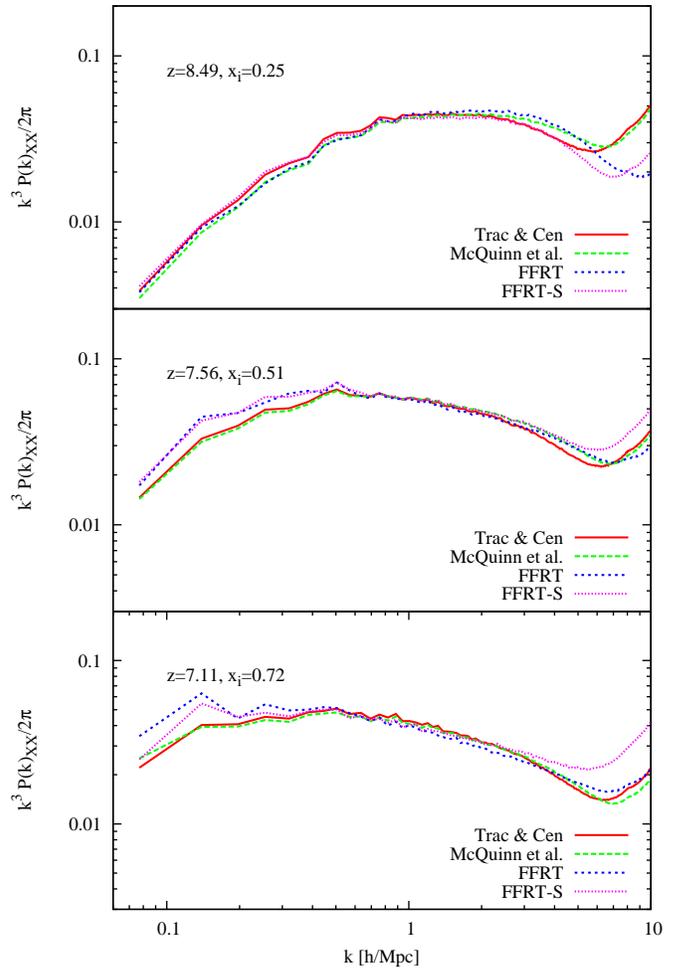}
\caption{Ionization field power spectra for the four schemes. As reionization progresses and bubbles grow and merge, the power shifts to larger scales. The semi-numeric schemes have more power on large scales due to their larger and more connected bubbles. The small scale upturn in power at $k \gtrsim 8$~h/Mpc is likely due to shot noise that is inherent in all of the schemes.}
\label{fig:pk_xi}
\ec
\end{figure}

\begin{figure}[t]
\bc
\includegraphics[width=9cm]{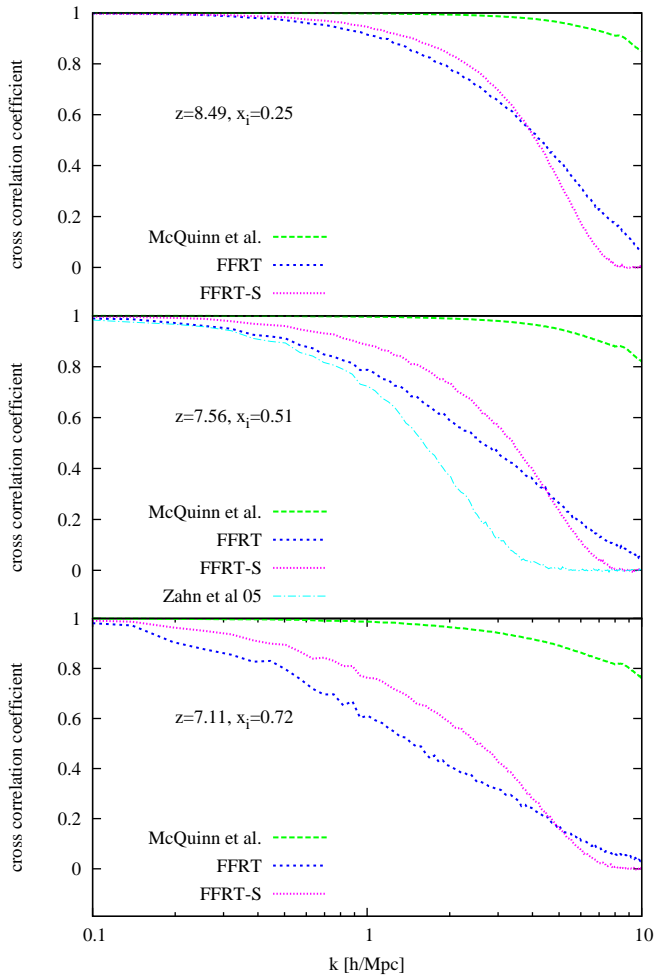}
\caption{Cross correlation coefficients between the \cite{TracCen2007} ionization fields and those from the other schemes. The two RT simulations show very good agreement on all scales at all times. Of the semi-numeric schemes, the FFRT-S comes closest to matching the RT results. For comparison, the \citet[][cyan dot-dashed]{Zahn2005} scheme is shown only in the middle panel. It shows poorest correlation owing to usage of the un-evolved density field and poor tracing of the density by the spherical top-hat filter.}
\label{fig:ccxi}
\ec
\end{figure}

Another measure of reionization morphology is the auto power spectrum of ionized regions $P_{\rm XX}$, which 
is relevant because it is directly related to the 21cm power spectrum (see Equation \ref{eq:pk21}). Figure \ref{fig:pk_xi} shows the evolution of the ionization power spectra. As previously noted \citep[e.g.][]{McQuinn2007b, Zahn2007}, the power shifts to larger scales as the bubbles grow and combine. Furthermore, the power spectrum is relatively flat as a function of scale during the final half of reionization, though semi-numeric studies predict a drop in power on scales larger than these simulation boxes (Mesinger \& Furlanetto 2007).  While all of our schemes show these trends, it is interesting to note that they are not as obviously present in the RT simulations of \citet{Iliev2006a}.

The small differences in $P_{\rm XX}$ between the four schemes are consistent with the differences seen previously in the bubble size distributions. At later times, the semi-numeric models have $\sim 30\%$ more large-scale power than the RT simulations because of their larger and more connected bubbles. As a more stringent test of the agreement between the ionization fields, we consider the cross correlation coefficient $r_{\rm X_1X_2} \equiv P_{\rm X_1X_2}/\sqrt{P_{\rm X_1X_1}P_{\rm X_2X_2}}$, where $P_{\rm X_1X_2}$ is cross power spectrum of ionization fields ${\rm X_1}$ and ${\rm X_2}$. While the schemes may agree on $P_{\rm XX}$ by having similar distributions of bubble sizes, $r_{\rm X_1X_2}$ specifically tests whether the ionized cells are in the correct locations by comparing the phases of the Fourier modes of the transformed ionization fields.

Figure \ref{fig:ccxi} shows the cross correlation between the \citet{TracCen2007} ionization fields and those of \citet{McQuinn2007b}, FFRT, and FFRT-S. On all scales, the ionization structure in the RT simulations agrees well, with $r_{\rm X_1X_2} >0.8$  for all $k$ out to the Nyquist mode. The slightly weaker correlation on scales near the grid spacing are due to small differences in the locations and shapes of the ionization fronts. The semi-numeric schemes show progressively weaker correlation as reionization progresses, with $r_{\rm X_1X_2}>0.8$ at $k \lesssim 1~$h/Mpc when $x_i ~ 0.5$. On most scales, the FFRT scheme performs less well than the FFRT-S scheme, because the discrete nature of sources is not captured in this excursion-set based analytic scheme. Note the significant improvement in the cross correlation coefficient in the FFRT scheme compared to the original semi-numeric reionization scheme, discussed in \citet{Zahn2005} [and shown only in the middle panel].  This improvement can be attributed mostly to the usage of the true density field (rather than the linear field) to calculate the collapse fraction.

\begin{figure}[t]
\bc
\includegraphics[width=9cm]{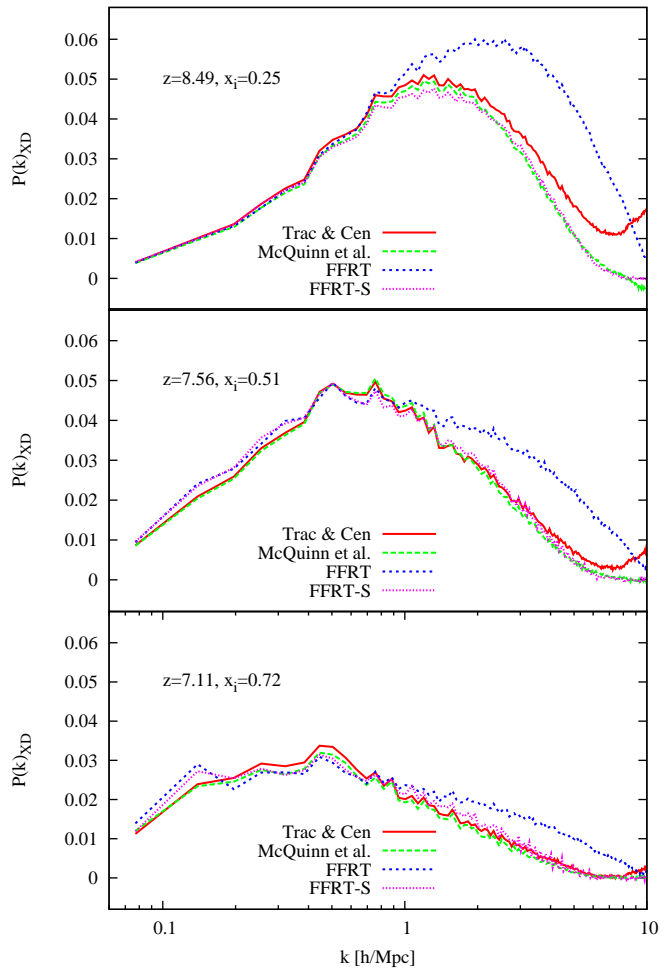}
\caption{Cross power spectra between the ionization and density fields for the four schemes. See Figure \ref{fig:pk_XDCCC} for the corresponding cross correlation coefficients. }
\label{fig:pk_XD}
\ec
\end{figure}

\begin{figure}[t]
\bc
\includegraphics[width=9cm]{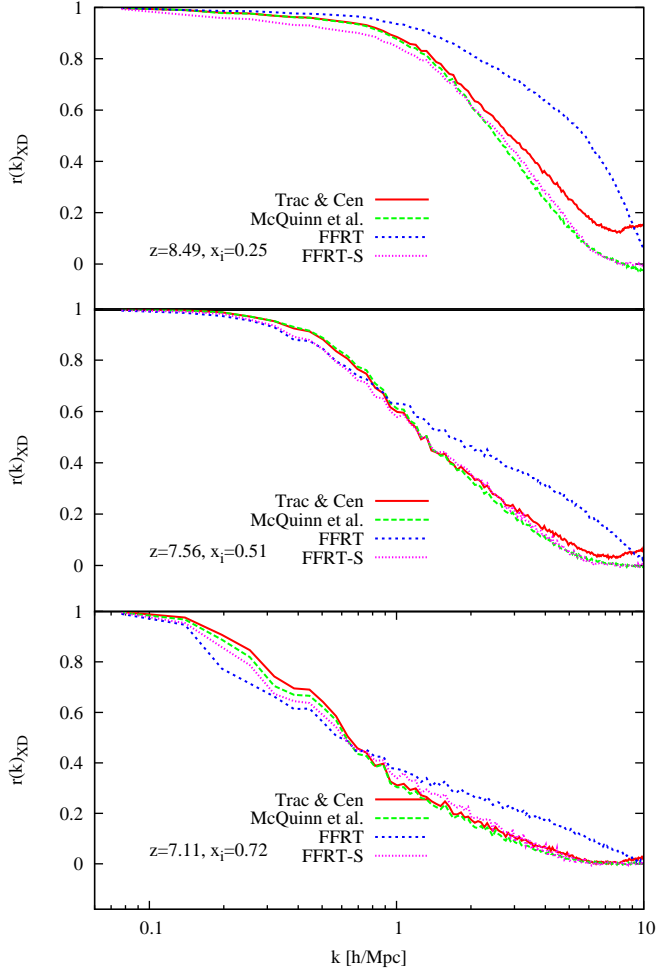}
\caption{Cross correlation coefficients between the ionization and density fields for the four schemes. The correlation is perfect on large scales, but it weakens with increasing $k$ because the sources (and sinks) are not perfectly correlated with the density field on small scales.}
\label{fig:pk_XDCCC}
\ec
\end{figure}

We also investigate how the bubbles are distributed with respect to the large-scale overdensities of dark matter and gas. As suggested already by many studies, large-scale overdense regions tend to be ionized earlier than large-scale underdense regions because they harbor more sources. The semi-numeric filtering techniques are motivated by this general description of the reionization process. In order to quantify the validity of this assumption, we calculate the cross power spectrum $P_{\rm XD}$ between the ionization and density fields, which is also directly related to the 21cm power spectrum (see Equation \ref{eq:pk21}). Figure \ref{fig:pk_XD} shows $P_{\rm XD}$ and Figure \ref{fig:pk_XDCCC} shows the cross correlation coefficient $r_{\rm XD} \equiv P_{\rm XD}/\sqrt{P_{\rm XX}P_{\rm DD}}$. In general, the ionization and density fields are perfectly correlated on large scales, but the correlation gets progressively weaker on smaller scales. This is expected as the sources (and sinks) are not perfectly correlated with the density field on small scales either.

The two RT simulations show good agreement overall in both $P_{\rm XD}$ and $r_{\rm XD}$. The \cite{TracCen2007} simulation appears to track the density field more closely, particularly for the ${x}_i =0.25$ case, with $r_{\rm XD}$ differing by $\sim0.1$ on scales near the grid spacing. The FRRT-S scheme compares very well with the two RT simulations, especially with the \citet{McQuinn2007b} results. There are some minor differences on large scales, but the good agreement overall validates the assumption discussed above. Not surprisingly, the FFRT scheme produces larger values of  $P_{\rm XD}$ and $r_{\rm XD}$ (particularly at $k > 1~$h/Mpc) compared to the other schemes because in this case, both the sources and ionization are determined by the evolved density field. Note that radiative transfer effects and poisson fluctuations in the source distribution  tend to decrease the cross correlation.

\section{Convergence in the $21$~cm Signal}\label{sec:21cm}

One of the main upcoming applications of reionization models will be to aid in the interpretation of upcoming high-redshift 21cm observations.
 Hence, we briefly explore the differences in the 21cm signal predicted by our various schemes.  We only study the power spectrum as this is a simple, often-studied statistic, which can encode information about the reionization state of the Universe (e.g., \citealt{Lidz2008})

Ignoring redshift-space distortions, the excess $21$~cm brightness temperature over that of the cosmic microwave background from emission at a redshift of $z$   
is (e.g., \citealt{ZaldarriagaFH2004}):
\beqa
\delta T
\, & \approx \, & 26 \, (1+\delta) \bar{x}_{HI}
\left( \frac{T_S - T_{\rm CMB}}{T_S} \right)
 \left( \frac{\Omega_b h^2}{0.022} \right) \nonumber \\
& & \times \left[ \left(\frac{0.15}{\Omega_m h^2} \right) \, \left(
\frac{1+z}{10} \right) \right]^{1/2} {\rm mK},
\label{eq:Tb}
\nonumber
\eeqa
where $\delta$ is the density contrast of gas, $T_S$ is the spin temperature, $T_{\rm CMB}$ is the temperature of the cosmic microwave background, and $\bar{x}_{HI} \equiv 1 - x_{i,V}$ is the average neutral fraction. 
The first-order expansion in ionization and density fluctuations for the $21$~cm power spectrum is
\beqa
P_{21 cm}(k)&=& \bar{T}_b^2 \left[P_{\rm XX} - 2\bar{x}_{HI}P_{\rm XD} + \bar{x}_{HI}^2 P_{\rm DD}\right],
\label{eq:pk21}
\eeqa
where $\bar{T}_b$ is the average brightness temperature in regions with $\bar{x}_{HI} = 1$, and $P$ is the power spectrum of the ionization (X) and density (D) fields.  Equation (\ref{eq:pk21}) ignores peculiar velocities and assumes $T_S \gg T_{\rm CMB}$.

\begin{figure}
\bc
\includegraphics[width=9cm]{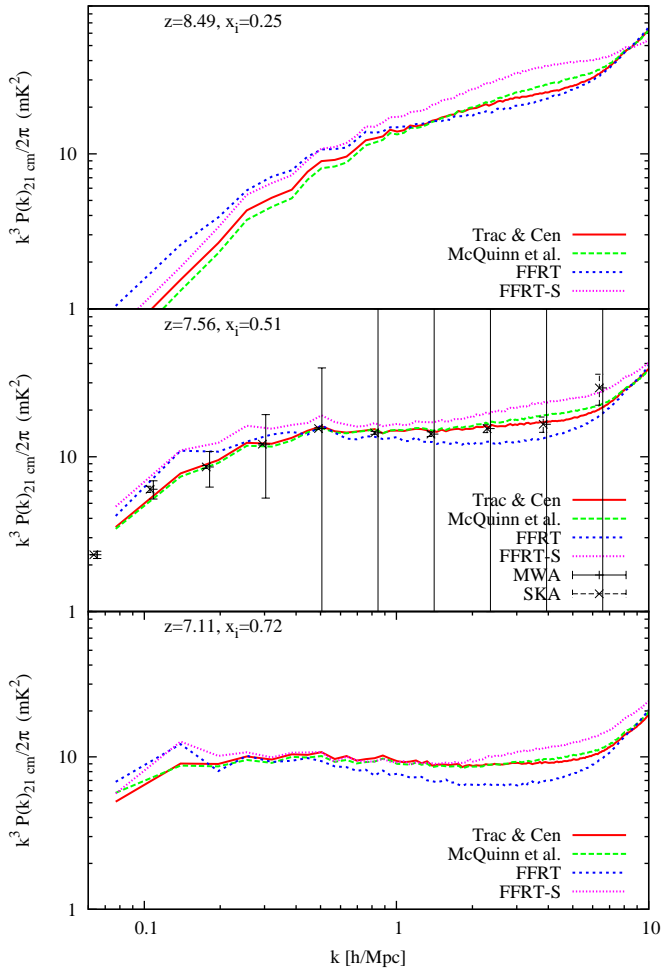}
\caption{Comparison of the 21 cm power spectrum in the four schemes, ignoring peculiar velocities.  At the intermediate ionization fraction we also compare the results of \cite{Zahn2005} and \cite{MesingerFurlanetto2007}, which agree less well with the simulations than the improved schemes proposed in this paper. Also included are the MWA and SKA forecasted errors.}
\label{fig:21cm_power}
\ec
\end{figure}

In Fig. \ref{fig:21cm_power}, we show the 21 cm power spectra computed using the four algorithms.  We assume $T_S \gg T_{\rm CMB}$, which previous studies have concluded should be accurate throughout most of reionization (e.g., \citealt{Furlanetto2006, PritchardFurlanetto2007}). We also neglect redshift-space distortions, which mainly affect smaller scales than the bubble scale during reionization \citep{McQuinn2006, MesingerFurlanetto2007}.  

As anticipated from the agreement between predictions for $P_{\rm XX}$  and $P_{\rm XD}$,  the prediction for $P_{21 cm}$ between the two radiative transfer simulations is in good agreement, differing by $\lesssim 10\%$ on all scales. The FFRT-S tends to have more power by $\sim30$ percent compared to the RT simulations, though this difference decreases as reionization progresses.
The FFRT algorithm has a more complicated behavior, tending to over-predict the large-scale power, and under-predict the small-scale power. The over-prediction of power on large scales is likely due to its over-connected HII regions. The under-prediction of power on small scales is due mainly to 
its overprediction of $P_{\rm XD}$ (see Figure \ref{fig:pk_XD} and eq.  \ref{eq:pk21}).
{\it Nevertheless, the 21 cm power spectra from all of the schemes agree with one another at the 10s of percent level.}

The error bars in the middle panel of Figure \ref{fig:21cm_power} are the forecasted precision that the MWA and the SKA can measure $P_{21 cm}$, assuming the configurations described in \citet{McQuinn2006}.  These errors assuming a $1000$~hr integration and a bandwidth of $6$~MHz.  The differences between the four schemes are smaller than the MWA 1-$\sigma$ errors at $k > 0.5$~h/Mpc. On larger scales, however, uncertainty in the semi-numerical modeling might bias reionization constraints within the errors of MWA, and especially with the SKA.
  However, it is debatable if we can accurately predict either the 21cm power spectrum or the relative differences among the ionization schemes on these large scales, given our limited box size.\footnote{Since all of the power spectra are computed from one realization of the density field, it is unclear how the mean signal and the differences between the models would change if the power spectra were computed from a much larger simulation box, or from an ensemble average (using different realizations of the initial conditions sampling the modes larger than our box size).}
Moreover, the relative differences among the ionization schemes are much smaller than the differences which can result from some astrophysical uncertainties during the EoR, such as the galaxy ionizing-photon luminosity function \citet{McQuinn2007b} and the number and distribution of absorption systems \citep{FurlanettoOh2005, ChoudhuryHR2009}.

\section{Conclusions}\label{sec:conclusions}

We have compared the results of different algorithms for computing the morphology of reionization.  Two of these algorithms were radiative transfer simulations, which made disparate approximations for propagating HII fronts from the sources, and the other two are less computationally expensive semi-numeric schemes, which use excursion-set formalism to identify ionized regions.  All of these schemes (or closely related variants) have been used in published studies of reionization (e.g., \citealt{Zahn2005,Zahn2007,McQuinn2007b,MesingerFurlanetto2007, McQuinn2007a, TracCen2007, MesingerFurlanetto2008a, TracCL2008}).  These algorithms were all compared on equal footing, using the same realization of the large scale matter distribution on which to generate the ionization field. 

The two radiative transfer schemes are in excellent agreement with each other (with the cross-correlation coefficient of the ionization fields given by $r_{\rm XX}>0.8$ for $k < 10$~h/Mpc) and in good agreement with
the analytic schemes ($r_{\rm XX}>0.6$ for $k < 1$~h/Mpc).  When used to predict the 21cm power spectra, one of the most important applications of reionization simulations in the near future, all ionization algorithms agree with one another at the 10s of percent level.
  The differences between these schemes are smaller than other uncertainties with modeling the EoR, such as the properties of the sources and sinks of ionizing photons.   This agreement suggests that the different approximations involved in the ray tracing algorithms are sensible and that semi-numerical schemes provide a numerically inexpensive, yet fairly accurate realization of the reionization process.  The speed and accuracy of these schemes can aid in efficient parameter space studies and in the interpretation of future observations.

\acknowledgments{OZ wishes to express special gratitude to the co-authors for being supportive in completing this work after he was injured in the fall of 2009. The authors wish to thank Adam Lidz, Steven Furlanetto, and Matias Zaldarriaga for helpful discussions. OZ is supported by an Inaugural Fellowship by the Berkeley Center for Cosmological Physics. AM is supported by NASA through Hubble Fellowship grant HST-HF-51245.01-A, awarded by the Space Telescope Science Institute, which is operated by the Association of Universities for Research in Astronomy, Inc., for NASA, under contract NAS 5-26555.  MM is supported by NASA through an Einstein Fellowship. HT is supported by an Institute for Theory and Computation Fellowship.  RC acknowledges support from NASA grant NNG06GI09G.}

\clearpage

\section{Appendix: Further Tests of the FFRT Scheme}

\subsection{Distributions of the Ionization Fraction}
\label{sec:xPDFs}

\begin{figure*}
\vspace{+0\baselineskip}
{
\includegraphics[width=0.5\textwidth]{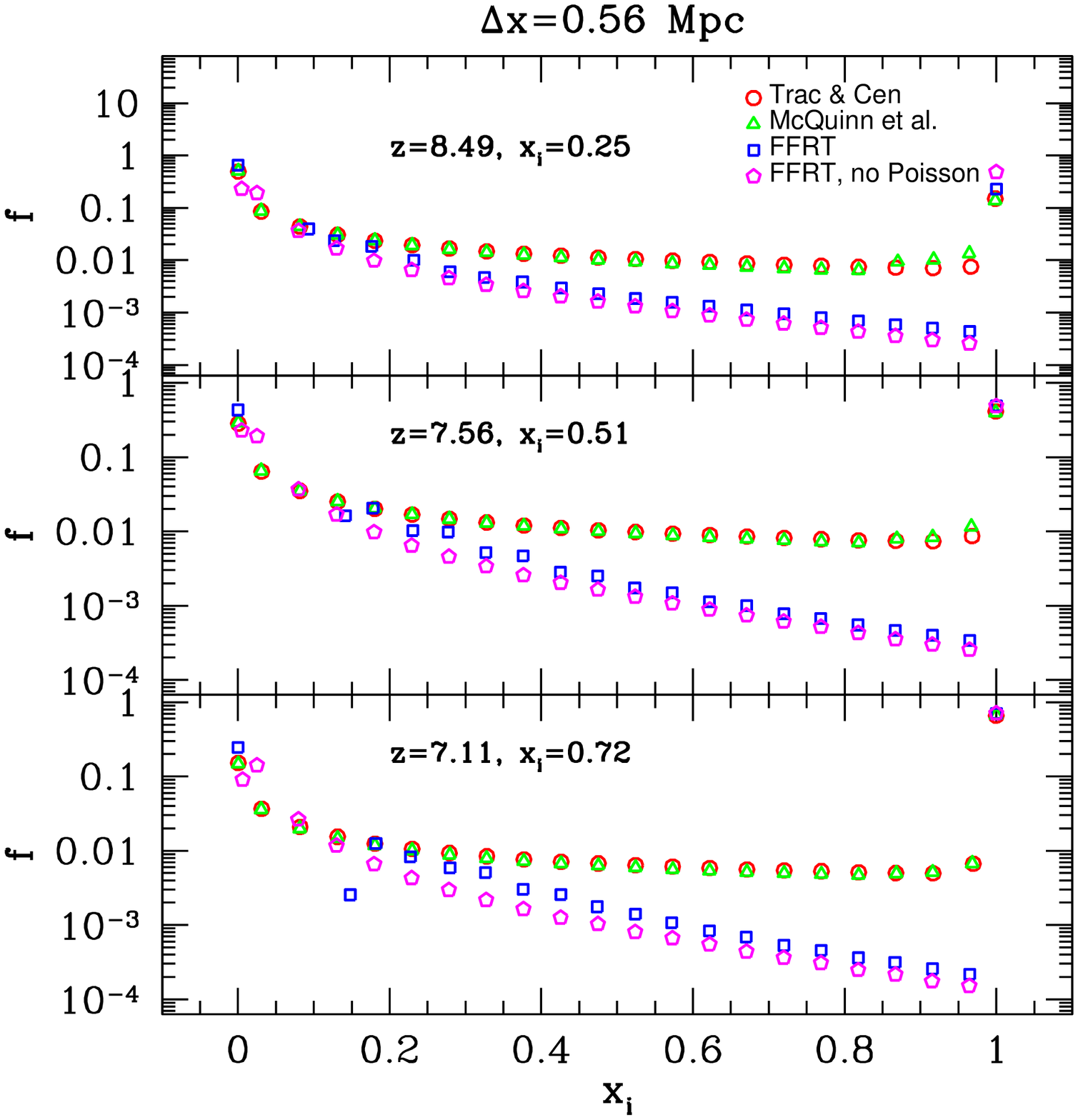}
\includegraphics[width=0.5\textwidth]{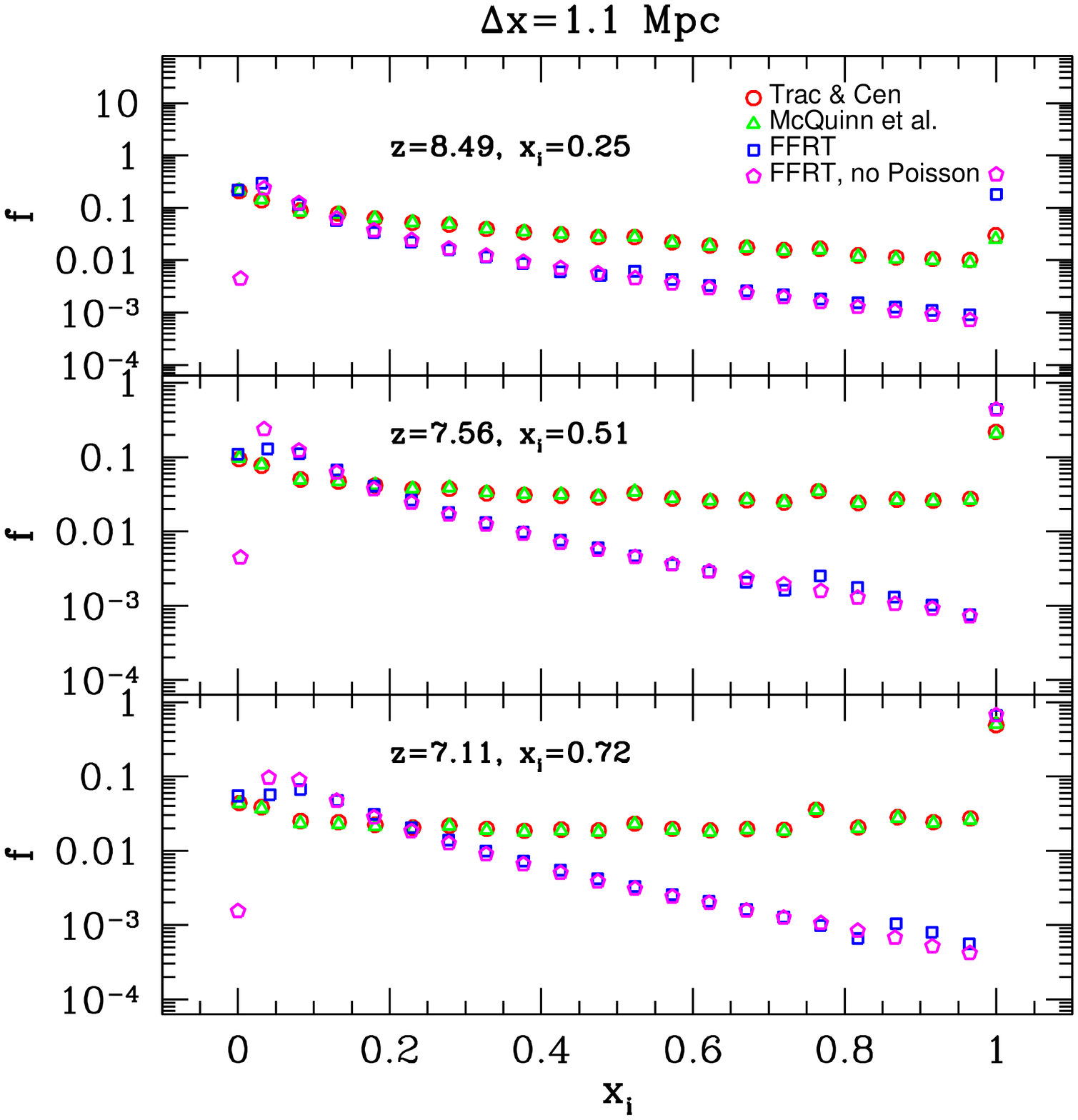}
}
\caption{
The fraction of simulation cells with a given ionization fraction, at $(z, x_{i,V})$= (8.49, 0.25), (7.56, 0.51), (7.11, 0.72), top to bottom.  Red circles, green triangles, blue squares, and magenta pentagons correspond to the ionization algorithms of \cite{TracCen2007}, \cite{McQuinn2007a}, FFRT, and FFRT without Poisson fluctuations in the halo number (see \S \ref{sec:seminumeric}), respectively.  The left panel was created from the 256$^3$ 
 grid (cell size $\Delta x=0.56$ Mpc)
 used throughout this paper, while the right panel was created from a 128$^3$ grid (cell size $\Delta x = 1.1$ Mpc).  Note that the distributions corresponding to the RT simulations in the right panel were computed from just the (boxcar filter) smoothed 256$^3$ grid.  However the FFRT ionization fields were generated directly from the low-resolution 128$^3$ root grid of the low-resolution density field (i.e., the $128^3$ FFRT ionization fields are not just smoothed versions of the 256$^3$ fields). \\ \\
\label{fig:PDFs}
}
\vspace{-1\baselineskip}
\end{figure*}

Here we investigate how well the analytic schemes perform at computing the level of partial ionization in simulation cells. This is where we expect the most disagreement between the RT and FFRT, as these cells contain the edges of HII regions/ionization fronts, where detailed radiative transfer effects are the most important.

We remind the reader that the FFRT scheme assigns partial ionization fractions based on the collapse fraction inside each cell.  Therefore, it takes into account the ionizing photons from sub-grid sources, which were not luminous enough to ionize the entire cell.  This is done in anticipation of the use of FFRT on very-large scales, with fairly large cells sizes (e.g., \citealt{Mesinger2009}).  As the cell size increases, fluctuations in the source number among neighboring cells decreases, and our approximation becomes increasingly accurate.  However, as the cell size becomes small ($M_{\rm cell} \sim M_{\rm min}$), partially ionized cells increasingly correspond to the edges of ionization fronts driven by external sources, instead of the sub-grid sources chewing away at their local HI as assumed by the FFRT scheme.  Here we wish to quantify these limits, checking how well the FFRT reproduces the distribution of partially ionized cells, given a certain cell size.

In Fig. \ref{fig:PDFs} we show the fraction of simulation cells with a given ionization fraction, at $(z, x_{i,V})$= (8.49, 0.25), (7.56, 0.51), (7.11, 0.72), top to bottom.  Red circles, green triangles, blue squares, and magenta pentagons correspond to the ionization algorithms of \cite{TracCen2007}, \cite{McQuinn2007a}, FFRT, and FFRT without Poisson fluctuations in the halo number (see \S \ref{sec:seminumeric}), respectively.  The left panel of Fig. \ref{fig:PDFs} shows the ionization fraction distributions for the 256$^3$ grid used throughout this paper. The two RT schemes agree extremely well, signaling again that these schemes have converged.   Note that on this grid scale, and assuming a relatively soft stellar ionization spectrum, the ionization field can indeed be well represented with a binary field.  The FFRT algorithm tends to under-predict the number of mostly-ionized cells.  This under-prediction is less severe in the early stages of reionization, when the surface area of ionization fronts is smaller.

Another interesting thing to note from the left panel is that our fiducial FFRT model which includes Poisson scatter in the sub-grid source number, has no partially ionized cells with ionization fractions less than $\sim 0.1$.  This is because a single source with our minimum halo mass $M_{\rm min}$ is luminous enough to ionize its host cell to a degree $\gsim 0.1$. On the other hand, if one ignores that sources are discrete and instead uses the mean conditional collapse fraction (magenta pentagons), the distribution of partial ionizations agrees fairly well with the RT in this regime.  This indicates that partial ionizations in the 256$^3$ grid with its $\Delta x = 0.56$ Mpc cells is dominated by unresolved ionization fronts driven by sources outside of each cell, as discussed above.  Hence, these cells are too small to justify the assumption of partial ionizations driven by sub-grid sources, and the good performance of the FFRT without Poisson scatter at $x_i \lsim 0.1$ is coincidental. 
Fortunately the ionization field on these scales is well approximated by a binary (fully neutral or fully ionized) field, and so correctly computing the partial ionization fraction is not very important for observables such as the 21cm power spectrum.

As discussed above, we expect our partial ionization scheme to be most effective when cell sizes are large, $M_{\rm cell} \gg M_{\rm min}$.  Unfortunately, we do not have RT fields for very large scales and cell sizes with which to check this assumption.  Instead we approximate larger cell sizes by smoothing the 256$^3$ density field down to 128$^3$, and using this lower-resolution grid as the {\it root grid} for the FFRT algorithm.  We compare the resulting distributions to the same RT fields (i.e., with the 256$^3$ root grid), but smoothed down to 128$^3$.  The resulting distributions are shown in the right panel of Fig. \ref{fig:PDFs}.  These distributions are not as smooth as the 256$^3$ ones, due to smaller number statistics.  However, we can note that the FFRT scheme does in fact perform better on these $\Delta x = 1.1$ Mpc cells.  This may or may not be a physical effect, due to the increasing relevance of sub-grid sources in setting the partial ionization fraction of larger cells.  We certainly expect agreement to improve with larger cell sizes.

Another interesting point evident from the left panel of Fig. \ref{fig:PDFs}, is that not including the discreteness of sources on this cell size results in a notable lack of fully neutral pixels.  This is to be expected since the mean value of the collapse fraction (which is the quantity directly computed in PS formalism) is never zero for $M_{\rm cell} > M_{\rm min}$.  Therefore, including the discrete nature of sources, such as with our Poisson scatter approach, is important in identifying pristine, fully-neutral regions on this scale.

\subsection{Impact of filter choice and cell flagging algorithm}
\label{sec:compare}

\begin{figure}
\bc
\includegraphics[width=9cm]{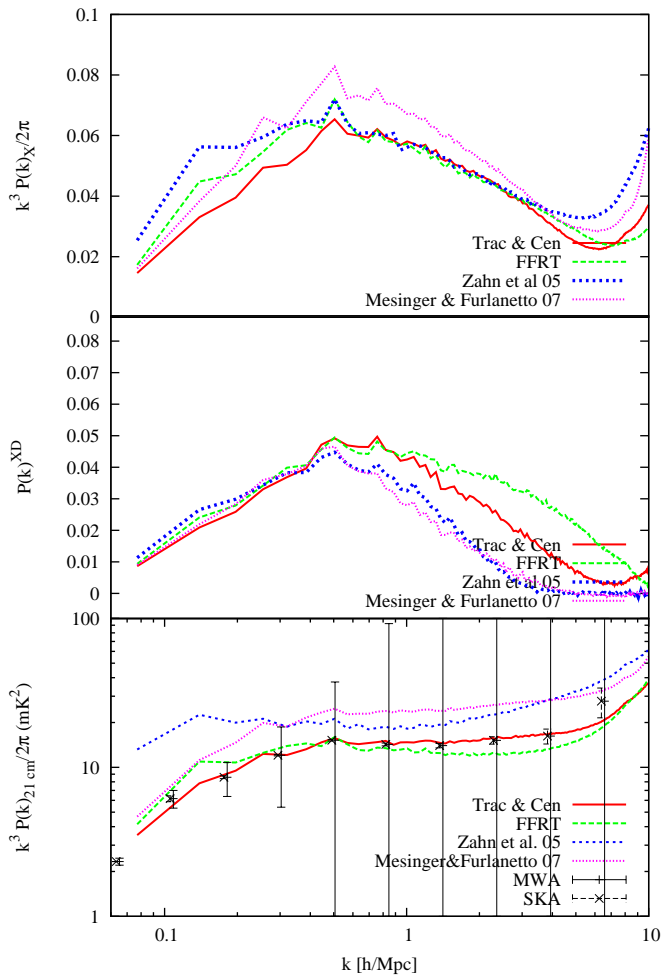}
\caption{Comparison of the ionization fraction power spectra ({\it top}), ionization-density cross power spectra ({\it middle}), and 21 cm power spectra ({\it bottom}) of the \cite{TracCen2007} RT ({\it solid red curves}) with several semi-numerical schemes. The FFRT corresponds to dashed green curves, Zahn et al. (2005) corresponds to short-dashed blue curves, and Mesinger \& Furlanetto (2007) corresponds to dotted magenta curves.}

\label{fig:compare_all_ffrt}
\ec
\end{figure}

Here we would like to explore how the filter choice and choice of cell flagging algorithm affect the resulting ionization fields.  Note that these are two of the main differences between our FFRT, FFRT-S algorithms, and those introduced in \cite{Zahn2005} and \cite{MesingerFurlanetto2007}. To do so we show the $P_{\rm XX}, P_{\rm X\delta}$, and $P_{\rm 21cm}$ power spectra of these models, together with the simulation that used the scheme of \cite{TracCen2007} for 
one
 redshift in Figure \ref{fig:compare_all_ffrt}.

These older schemes do not include partial ionizations, though this has a minimal effect on these scales as seen above.  Aside from this,
the \citet{Zahn2005} scheme is different from the FFRT mainly in using a top-hat instead of a sharp k-space filter and operating on the linear density field, instead of the evolved density field.  The \citet{MesingerFurlanetto2007} scheme is different from the FFRT-S mainly in painting the entire spherical region enclosed by the filter as ionized (``flagging entire sphere''), instead of just the central filter cell (``flagging center cell'').

All models do a decent job of predicting the ionization power spectra.  However, it can be seen from the middle panel that both of the older schemes under-predict the cross-correlation of the ionization and the density fields on moderate scales.  As shown in eq. (\ref{eq:pk21}), this underprediction results in an increase in 21cm power, as can be seen in the bottom panel.  The sharp k-space filtering (FFRT) reaches more small scale features in the density field, comparable to the RT algorithm, conversely tending to over-predict the ionization-density correlation $P_{\rm xd}$ on small scales, yielding a slight under-estimate of $P_{\rm 21cm}$ there.  Note that our FFRT-S scheme does very well in predicting this cross-correlation, suggesting that one has to use the center-flagging procedure {\t and} either use k-space filtering or use discrete sources not to get this underestimate.  Interestingly, the {\it shape} of the power spectrum is accurately predicted by the \citet{MesingerFurlanetto2007} algorithm.
 Overall the new semi-numeric schemes introduced here fare best in comparison to the radiative transfer simulation. 

To summarize, if one is interested in just the ionization fields, all models do a good job.  The ``flagging the entire sphere'' algorithm performs slightly better than ``flagging center cell'' in ``by-eye'' and cross-correlation comparisons with RT simulations.  The situation is more complicated if one is interested in the 21cm fields, as cross-correlations with the density field and higher-order terms become relevant \citep{Lidz2009}.  In this regime, ``flagging the center cell'' (instead of the entire sphere) seems to predict a better cross-correlation of the ionization and density fields, and is a simpler and faster algorithm.  To get accurate 21cm power spectra, one should additionally either use a sharp k-space filter on the evolved density field (FFRT) or use discrete sources (FFRT-S).

\bibliographystyle{apj}
\bibliography{reion}

\end{document}